\documentclass[amssymb, preprintnumbers, showpacs showkeys,aps,prl,superscriptaddress,twocolumn,nobibnotes]{revtex4-1}

\usepackage{color} 
\usepackage{xcolor} 
\usepackage{hyperref}
\usepackage{slashed}
\usepackage{physics}
\usepackage{graphicx}
\usepackage{amsmath}
\usepackage{latexsym}
\usepackage{epstopdf}
\usepackage{amsmath}
\usepackage{enumitem}
\usepackage{amssymb,amsmath}
\usepackage{multirow}
\usepackage{mathtools}
\usepackage{bbm}
\usepackage{bm}
\usepackage{amsfonts}
\usepackage{amssymb}
\usepackage{calligra}
\usepackage{calrsfs}
\usepackage{makecell}
\usepackage[normalem]{ulem}
\usepackage[USenglish,american]{babel}
\usepackage{subfigure}

\usepackage{dsfont}

\expandafter\ifx\csname package@font\endcsname\relax\else
 \expandafter\expandafter
 \expandafter\usepackage
 \expandafter\expandafter
 \expandafter{\csname package@font\endcsname}%
\fi
\begin{document}

\title{Low energy excitations in bosonic quantum quasicrystals}

\author{A. Mendoza-Coto}%
\email{alejandro.mendoza@ufsc.br}
\affiliation{Departamento de F\'\i sica, Universidade Federal de Santa Catarina, 88040-900 Florian\'opolis, Brazil}%
\affiliation{Max Planck Institute for the Physics of Complex Systems, Nothnitzerstr. 38, 01187 Dresden, Germany}

\author{M. Bonifacio}%
\affiliation{Max Planck Institute for the Physics of Complex Systems, Nothnitzerstr. 38, 01187 Dresden, Germany}%

\author{F. Piazza}%
\affiliation{Max Planck Institute for the Physics of Complex Systems, Nothnitzerstr. 38, 01187 Dresden, Germany}
\affiliation{Theoretical Physics III, Center for Electronic Correlations and Magnetism,
Institute of Physics, University of Augsburg, 86135 Augsburg, Germany}

\begin{abstract}
   We present the first principles construction of the low-energy effective action for bosonic self-organized quantum quasicrystals. Our generalized elasticity approach retains the appropriate number of phase- and corresponding conjugate density- degrees-of-freedom required for a proper description of the Goldstone modes. For the dodecagonal and decagonal quasicrystal structures we obtain collective longitudinal and transversal excitations with an isotropic speed of sound. Meanwhile, for the octagonal structure, the coupling between phononic and phasonic degrees of freedom leads in turn to hybridization of the latter with the condensate sound mode, producing collective excitations with a longitudinal and transversal component, and an anisotropic speed of sound. Finally, we discuss the fate of each excitation mode at the low and high density phase transitions limiting the quantum quasicrystal phase.

\end{abstract}

\maketitle

\textit{Introduction.}--Quasicrystals (QCs) are an exotic state of matter where crystalline and disordered behaviour meet. They retain long range orientational order but lack any discrete translational invariance~\cite{Ja1997,She1951,Levine1984}.  Although they were initially discovered in metallic alloys~\cite{An2008,Lu2009}, it is known today that this kind of exotic state can be stabilized in many different contexts~\cite{Li1997,Freedman2007,Ba2011,Zu2017}, as long as an appropriate effective competition between two or more characteristic length scales is present in the system~\cite{Ba2014,Pupillo2020,Abreu2022,MeZaDi2022}. One of the most intriguing versions of quasi crystalline systems  are the so-called quantum quasicrystals~\cite{Gopa2013,Mivehvar2019,MeZaDi2022,Gross2023,Ciardi2023,Gautier2021,Zhu2023}, which appear as ground states of quantum many-body Hamiltonians. Such is the case of QCs produced by spin-orbit interactions in dipolar bosons~\cite{Gopa2013}, quasicrystalline phases induced in ultracold bosonic gases by quasiperiodic optical traps~\cite{Ciardi2023,Gautier2021,Zhu2023,Viebahn2019,Sbroscia2020} and more recently, self-organized quasicrystals in Bose-Einstein condensates produced by cavity mediated interactions~\cite{Mivehvar2021,Mivehvar2019}.
In this Letter, we present the first principles construction of the low energy elastic theory of bosonic quantum quasicrystals, having supersolids as a particular case. Our calculations maintain full connection with the microscopic properties of the system allowing us to make qualitative and quantitative predictions regarding the hybridization of the low momentum gapless energy modes, the corresponding propagation speed and the behavior of these excitations at the phase boundaries of the  quasicrystal phase. As particular examples we consider the three most relevant quasiperiodic structures in two dimensions, the octagonal, the decagonal and the dodecagonal quasicrystals. To our knowledge, this is the first microscopic elastic theory of bosonic quantum quasicrystals able to corroborate the symmetry arguments regarding the existence of five gapless excitation modes for these systems. Our results show that the hybridization process between phonons, phasons and the longitudinal condensate sound mode qualitatively depends on the kind of quasicrystal structure.
\begin{figure*}[t]
\centering
\includegraphics[width=\textwidth]{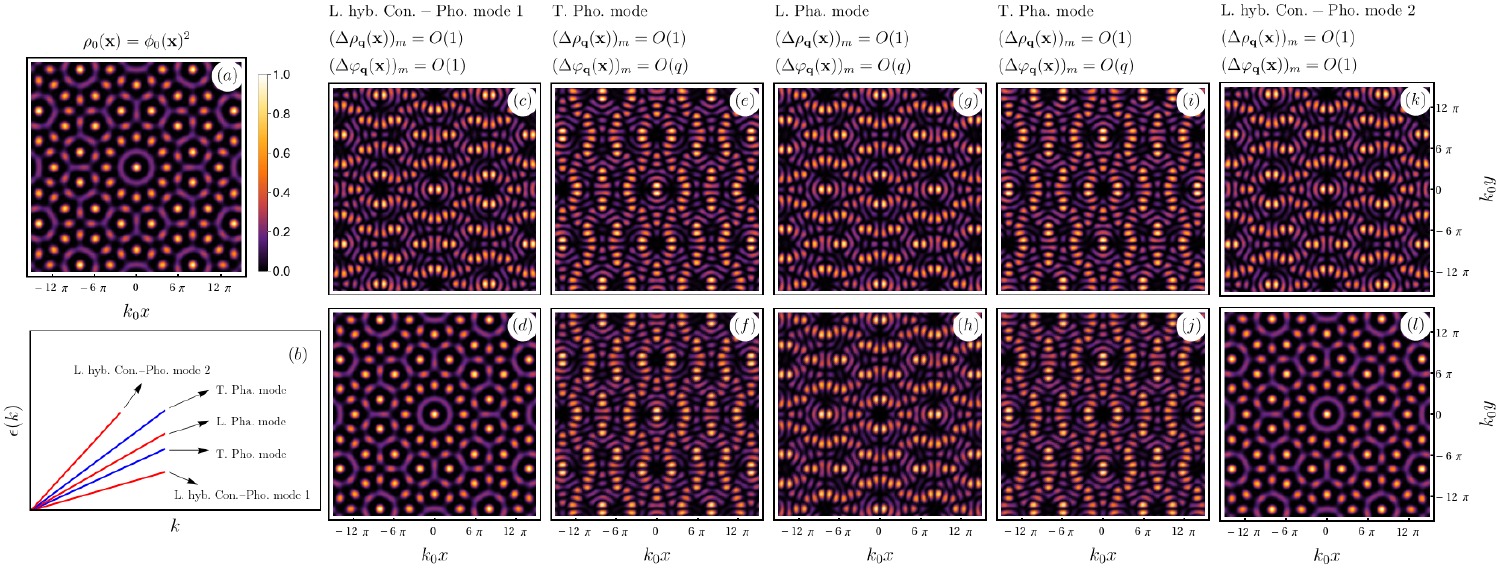}
        \caption{Typical density pattern \textbf{(a)} and low-energy excitations of a dodecagonal quantum quasicrystal (QQC) in the low momentum regime \textbf{(b-l)}. The color scale included in \textbf{(a)} applies to all figures provided each profile is normalized by the maximum of the corresponding quantity. \textbf{(b)} Dispersion relation $\epsilon(k)$ for the gapless excitation modes of a dodecagonal QQC. \textbf{(c-l)} Density fluctuations $\Delta \rho_\textbf{q}(\textbf{x})$ (first row) and phase fluctuations $\Delta \varphi_\textbf{q}(\textbf{x})$ (second row) in the low momentum regime for each gapless excitation mode as they appear in \textbf{(b)} from left to right. The scaling behaviour of the density and phase fluctuations for each mode is given at the top of each column. The momentum of the excitation $\textbf{q}$ is chosen along the "$x$" axis in all cases, see Ref.~\cite{sm} for detailed information. 
    }
    \label{fig1}
\end{figure*}
\\ \textit{Model and Method.}--We consider a 2D bosonic gas at zero temperature, interacting through an isotropic non-local two-body interaction of the form $U v(r)$, where $U$ represents a dimensionless parameter characterizing the intensity of the non-local interaction with respect to the kinetic energy. The nature of the two-body interaction between particles is such that the system hosts a quasicrystalline pattern at high enough particle densities in the ground state. As discussed in the literature~\cite{MeZaDi2022,Gross2023,Heinonen2019}, the key ingredient for the stabilization of such phases is the presence of competing minima in the interaction potential $\hat{v}(k)$. In dimensionless units, the path-integral formulation of the grand canonical partition function of the system $\mathcal{Z}=\int D\phi D\phi^*e^{-S[\phi,\phi^*]}$~\cite{Stoof2009} is characterized by the action: 
\begin{eqnarray}
\nonumber
S[\phi,\phi^*]&&=\int_0^\infty d\tau\int d^2x \ \phi(\textbf{x},\tau)^*\left(\partial_{\tau}-\frac{1}{2}\nabla^2-\mu\right.\\
&&\left.+\frac{U}{2}\int d^2x'v(\textbf{x}-\textbf{x}')\vert\phi(\textbf{x}',\tau)\vert^2 \right)\phi(\textbf{x},\tau).
\label{original_action}
\end{eqnarray}\\
To access the low-energy properties of the system we should allow long-wavelength fluctuations of the global phase, as well as of the phase of the modulated pattern of the quasicrystal ground-state wave-function which can be expanded in a Fourier basis as $\phi_0(\textbf{x})=\sqrt{\rho}\sum_{\textbf{n}}c_{\textbf{n}}\exp(i\textbf{G}_{\textbf{n}}\cdot\textbf{x})$. Here $\rho$ stand for the average particle density of the system and $c_\textbf{n}$ and $\textbf{G}_\textbf{n}$ corresponds to the Fourier amplitudes and wave vectors of the pattern respectively. We thus propose a perturbed ground state wave function of the form:  
\begin{eqnarray}
    &&\nonumber\phi(\textbf{x},\tau)=\sqrt{\rho}c_0\sqrt{1+\frac{\delta n_0(\textbf{x},\tau)}{\rho c_0^2}}\exp(i\theta(\textbf{x},\tau))\\ \nonumber
    &+&\sum_{\textbf{n}\neq0}\sqrt{\rho}c_\textbf{n}\sqrt{1+\frac{\textbf{G}_{\textbf{n}}\cdot\bm{\Pi}(\textbf{x},\tau)}{Z}+\frac{\textbf{G}_{\textbf{n},\perp}\cdot\bm{\Pi}_{\perp}(\textbf{x},\tau)}{Z_{\perp}}}\\ \nonumber
    &&\exp\left(i \textbf{G}_{\textbf{n}}\cdot\textbf{x}+i \textbf{G}_{\textbf{n}}\cdot\textbf{u}(\textbf{x},\tau)
    + i \textbf{G}_{\textbf{n},\perp}\cdot\textbf{w}(\textbf{x},\tau)\right.\\
    &+&\left.i\theta(\textbf{x},\tau)\right),
    \label{Eqflut}
\end{eqnarray}
where the fluctuation fields $\textbf{u}(\textbf{x},\tau)$ and $\textbf{w}(\textbf{x},\tau)$ represent the phonon and phason fields introduced as in the classical case~\cite{Levine1985,Socolar1989,Ding1993}. Moreover, due to the quantum nature of the system we also have to admit fluctuations in the canonical-conjugate density fields associated to these phase variables, analogously to what was proposed by Pretko et al.~\cite{Pretko2018} for supersolids. Accordingly, we have included the fields $\bm{\Pi}(\textbf{x},\tau)$ and $\bm{\Pi}_\perp(\textbf{x},\tau)$ as conjugate to $\textbf{u}(\textbf{x},\tau)$ and $\textbf{w}(\textbf{x},\tau)$, respectively. Finally, we added the overall phase fluctuation field $\theta(\textbf{x},\tau)$ of the wave function and its canonically conjugate zero momentum density fluctuation $\delta n_0(\textbf{x},\tau)$. It is understood that all fluctuation fields are slowly varying functions in space and imaginary time, with zero mean over the size of the system and, as expected, in the absence of fluctuations our ansatz for $\phi(\textbf{x},\tau)$ reduces to the ground state wave function $\phi_0(\textbf{x})$. Furthermore, the set of vectors $\{\textbf{G}_{\textbf{n},\perp}\}$ is constructed from a permutation of the set $\{\textbf{G}_{\textbf{n}}\}$, such that the extended basis $\tilde{\textbf{G}}_{\textbf{n}}=\textbf{G}_{\textbf{n}}\oplus\textbf{G}_{\textbf{n},\perp}$ is orthogonal in the sense $\sum_{\textbf{n}}\tilde{\textbf{G}}_{\textbf{n},\alpha}\tilde{\textbf{G}}_{\textbf{n},\beta}\propto\delta_{\alpha,\beta}$~\cite{Levine1985,Socolar1989,Ding1993}.

It is simple to show that the proposed ansatz satisfies the normalization condition of the boson field  $\int_A d^2x\left<\vert\phi(\textbf{x},\tau)\vert^2\right>=N=\rho A$, as long as $\sum_{\textbf{n}}c_\textbf{n}^2=1$. Such normalization constraint allows the calculation of the chemical potential $\mu$ for a system with a given particle density $\rho$. Lastly, the free parameters $Z$ and $Z_\perp$ in Eq. (\ref{Eqflut}) are chosen to enforce the expected relation between canonically conjugate fields in the low energy effective action of the system~\cite{sm}. By replacing the above ansatz for $\phi(\textbf{x},\tau)$ in Eq. \eqref{original_action} and expanding up to quadratic order in the fluctuation fields we obtain the following long-wavelength effective action:
 \begin{align}
    \nonumber
\delta S&=\int d\tau d^2x \left(i \delta n_0(\textbf{x},\tau)\partial_\tau\theta(\textbf{x},\tau)+i \bm{\Pi}(\textbf{x},\tau)\cdot\partial_\tau\textbf{u}(\textbf{x},\tau)\right.\\ \nonumber
&+\left.i \bm{\Pi}_\perp(\textbf{x},\tau)\cdot\partial_\tau\textbf{w}(\textbf{x},\tau)\right)+ \int d\tau d^2x\ \frac{1}{2}\left[\rho(\bm{\nabla}\theta)^2\right.\\ \nonumber
&+\gamma_0\delta n_0(\textbf{x},\tau)^2+\gamma\bm{\Pi}(\textbf{x},\tau)^2+\gamma_\perp\bm{\Pi}_\perp(\textbf{x},\tau)^2\\&+2\bm{\Pi}\cdot\bm{\nabla}\theta+\left.\textbf{E}^T~\tilde{C}~\textbf{E}\right],
\label{action}
\end{align}
where $\textbf{E}$ stands for the extended phonon-phason strain field defined as $\textbf{E}=\{\partial_xu_x,\partial_yu_y,(\partial_xu_y+\partial_yu_x)/2,(\partial_yu_x+\partial_xu_y)/2,\partial_xw_x,\partial_yw_y,\partial_yw_x,\partial_xw_y\}$ and $\tilde{C}$ represents the phonon-phason elastic tensor. Closed formulae for the elements of this tensor as well as for the $\gamma$'s couplings in terms of the ground state wave function are provided in the Supp. Matt.~\cite{sm}. The expression in Eq. (\ref{action}) for the low-energy effective action of a bosonic quasicrystal is one of the main results of this work. It allows qualitative and quantitative predictions since all couplings are known in terms of the ground state wave function of the system. Moreover, it generalizes the well-known density-phase action of a homogeneous condensate~\cite{Stoof2009}, not only to the present case of quantum quasicrystals, but also to supersolids if we eliminate the phasonic degrees of freedom \cite{kunimi2012,Ye2008,Pomeau_Rica_1,Pomeau_Rica_2,Pomeau_Rica_3,Heinonen2019}.

\textit{Elementary excitations for the dodecagonal QC}-- The effective action in Eq. \eqref{action} can be recast as $\delta S=\frac{1}{2}\int \frac{d\omega}{(2\pi)}\frac{d^2q}{(2\pi)^2}\bm{\hat{\eta}}(\textbf{q},\omega)^\dagger \textbf{M}(\textbf{q},\omega)\bm{\hat{\eta}}(\textbf{q},\omega)$, where $\bm{\hat{\eta}}(\textbf{q},\omega)=(\delta \hat{n}_0(\textbf{q},\omega),
\hat{\theta}(\textbf{q},\omega),
  \hat{\Pi}_x(\textbf{q},\omega),
  \hat{u}_x(\textbf{q},\omega),
  \hat{\Pi}_y(\textbf{q},\omega),
  \hat{u}_y(\textbf{q},\omega),$ $
  \hat{\Pi}_{x,\perp}(\textbf{q},\omega),
  \hat{w}_x(\textbf{q},\omega),
  \hat{\Pi}_{y,\perp}(\textbf{q},\omega),$ $
  \hat{w}_y(\textbf{q},\omega))$, the energy-momentum dispersion relations $\epsilon(\textbf{q})=\omega(\textbf{q})$ can be computed by solving the equation $\det\left[\textbf{M}(\textbf{q},i\omega(\textbf{q}))\right]=0$~\cite{Stoof2009}. The dodecagonal quasicrystal (Fig. \ref{fig1}(a)) is the simplest case due to the absence of the coupling between phonons and phasons. The latter are deformations of the density pattern which are absent in ordinary crystals. We obtain analytical expressions for the dispersion relation of the five gapless modes~\cite{sm}. In this case there are three longitudinal modes and two transversal modes, as schematically represented in Fig. \ref{fig1}(b). Two of the longitudinal modes result from the hybridization of the condensate sound mode (corresponding to a modulation of the overall phase $\theta$) with the longitudinal phonon mode, while the transversal phonon as well as the phason mode remain decoupled.

To deepen our understanding of the excitation modes we calculate the eigenvectors $\bm{\eta}_j(\textbf{q})$ using the charateristic equation $\textbf{M}(\textbf{q},i \omega_j(\textbf{q}))\bm{\hat{\eta}}_j(\textbf{q})=\textbf{0}$ for each excitation mode. Each eigenvector corresponds to a fluctuation field in real space and time given by $\bm{\eta}_j(\textbf{x},t)=\mathrm{Re}\left[\bm{\hat{\eta}}_j(\textbf{q})\exp(i(\textbf{q}\cdot\textbf{x}-\omega_j(\textbf{q})t))\right]$. By expanding Eq. \eqref{Eqflut} to linear order in the fluctuation fields we obtain a general expression for the correction to the ground state wave function: $\delta\psi_{j,\textbf{q}}=[u_{j,\textbf{q}}(\textbf{r})\exp(i(\textbf{q}\cdot\textbf{r}-\omega_j(\textbf{q})t))-v_{j,\textbf{q}}(\textbf{r})^*\exp(-i(\textbf{q}\cdot\textbf{r}-\omega_j(\textbf{q})t))]$, allowing us to identify the Bogoliubov modes $u_{j,\textbf{q}}(\textbf{r})$ and $v_{j,\textbf{q}}(\textbf{r})$~\cite{sm}. 
The local phase and density fluctuations can be then written as $\Delta\rho_{j,\textbf{q}}=\vert u_{j,\textbf{q}}(\textbf{r})-v_{j,\textbf{q}}(\textbf{r})\vert^2$ and $\Delta\varphi_{j,\textbf{q}}=\vert u_{j,\textbf{q}}(\textbf{r})+v_{j,\textbf{q}}(\textbf{r})\vert^2$~\cite{Macri2013, Wu1996}. In Fig. \ref{fig1}(\textbf{c}-\textbf{l}) we show the typical behavior of these quantities. Each column corresponds to a mode in Fig. \ref{fig1}(\textbf{b}) in increasing order of excitations energy. In agreement with previous observations for supersolids~\cite{Macri2013}, the hybridized  longitudinal modes present significant ($O(1)$) phase and density fluctuations, while the transversal phonon mode presents significant density fluctuations and  $O(q)$ phase fluctuations. On the other hand, while showing the same scaling ($O(1)$ vs $O(q)$) of the transversal phonon modes, phasons feature density and phase fluctuations patterns which presents a different behavior around the center of the quasicrystal structure~\footnote{The center of the quasicrystal here is being understood as the special point where all reflections symmetry planes of the quasicrystal pattern intersects. Given the form of our anzats for $\phi_0(\textbf{x})$ it coincides with the origin of our coordinates system.} compare e.g. panel (e) and (i) of Fig.\ref{fig1}.  Mathematically, this is ultimately a consequence of the phason's property of producing a redistribution of the QC structure destroying the center of the quasicrystal without altering the average properties of the system~\cite{Li2011}. Finally, we would like to remark that differently from the case of supersolids, here all density and phase fluctuations patterns are aperiodic.

\begin{figure*}[t]
    \centering
\includegraphics[width=0.51\linewidth]{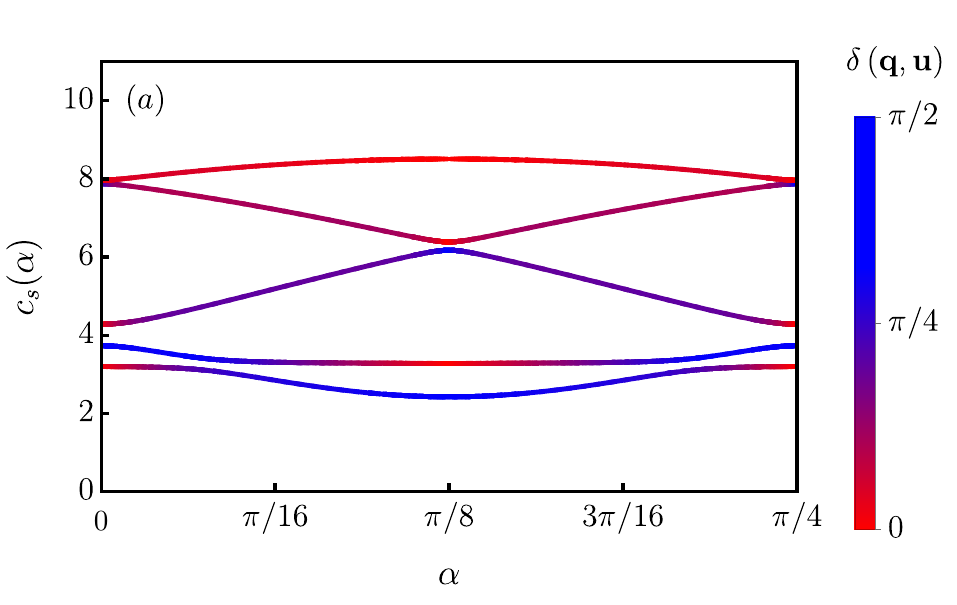}  
\includegraphics[width=0.45\linewidth]{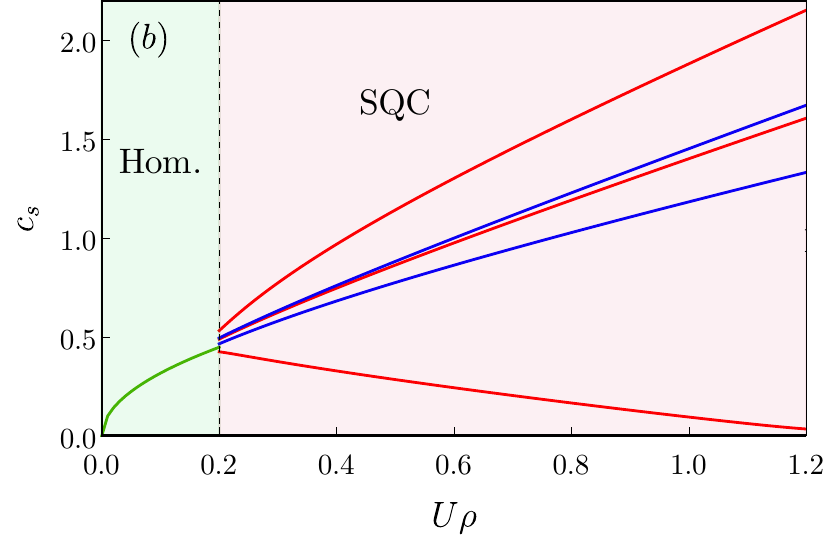}
\caption{Typical behavior of the propagation speed of the gapless modes, (a) as a function of the orientation $\alpha$ of the momentum of the excitation $\textbf{q}$ in a octagonal QC and (b) as a function of $U\rho$ in a decagonal QC. The red to blue gradient of color is used to indicate the transversal to longitudinal character of a given excitation mode using the angle ($\delta (\textbf{q},\textbf{u})$) of the field $\textbf{u}$ respect to $\textbf{q}$ as a measure of this property.}
    \label{fig2}
\end{figure*}

\textit{Excitations in decagonal and octagonal QCs.}--In the case of ten- and eight-fold quasicrystals, the presence of the phonon-phason coupling leads to a characteristic equation $\det\left[\textbf{M}(\textbf{q},i\omega(\textbf{q}))\right]=0$ with convoluted analytical solutions. However, some general properties can be obtained from the numerical solution of this problem. In the case of the decagonal quasicrystal, the longitudinal and transversal modes separately hybridize among themselves, leading to two groups of three and two modes, respectively, all with isotropic dispersion relations. In contrast, for the octagonal quasicrystal, the structure of the elasticity action characterized by an anisotropic phason contribution and a finite phonon-phason coupling produces anisotropic dispersion relations $\epsilon(\textbf{q})$ with propagation speed $c_s$ that depend on the orientation of the momentum $\textbf{q}$ when $q\rightarrow0$.  Additionally, unless the momentum of the excitation $\textbf{q}$ is oriented along one of the symmetry axes of the quasicrystal, every mode has a longitudinal and a transversal component. As an example, we show in Fig. \ref{fig2}(a) the anisotropic behavior of the propagation speed and hybribization of modes for an octagonal QC. The numerical evaluation was carried out considering values of the couplings for which the system displays strong anisotropy and  hybridization effects~\cite{sm}. The red to blue color gradient indicates the longitudinal and transversal character of each mode, using the angular orientation of the field $\textbf{u}$ as a measure of this property. Interestingly, the propagation of excitations along non-equivalent symmetry axes of the octagonal quasicrystal produces an exchange of the longitudinal and transversal character for some modes. Moreover, the evaluation of the local density and phase fluctuations profile in this case shows that the scaling of the fluctuations when the propagation direction of the excitation do not coincide with a symmetry axes of the quasicrystal are all $O(1)$. Meanwhile, in the opposite case, longitudinal modes have density and phase fluctuations of amplitude $O(1)$ and transversal modes with density fluctuations $O(1)$ and phase fluctuations $O(q)$. The aperiodic patterns of phase and density fluctuations for each scenario discussed can be found in \cite{sm}.

\textit{Instabilities of the quantum quasicrystal phase.}-- A first analysis of the instabilities can be based on the behaviour of the propagation speed of the gapless modes. With this aim we consider the case of the decagonal QC structure and pursue the calculation of the couplings in the action \eqref{action}, considering small amplitudes of the QC modulation~\cite{sm}. 
In order to make a numerical evaluation possible without defining a particular form of the interaction potential, we assume that the QC amplitude is a concave increasing  function of $\rho U$~\footnote{Mathematically this means $\frac{\partial c_1(U\rho)}{\partial (U\rho)}>0$ and $\frac{\partial^2 c_1(U\rho)}{\partial (U\rho)^2}<0$, where $c_1(U\rho)$ stand for the modulation amplitude of $\phi_0$.}. Although this is the expected behavior of this quantity, such ansatz does not intend to capture the actual quantitative behavior of a certain model~\cite{sm}. The results for the propagation speed of each mode are shown in Fig. \ref{fig2}(b) as a functon of $\rho U$. One recognizes a discontinuous behavior of the propagation speed at the homogeneous-superfluid-to-quantum-quasicrystal phase transition, analogous to previous observations in supersolids~\cite{Macri2013,Blakie2023, Heinonen2019,kunimi2012}, and a continuous behavior at the quantum-to-classical quasicrystal transition at high enough densities where the propagation speed of the lowest mode vanishes~{\cite{Saccani2012},\footnote{Although Mean Field treatments of our action consistent with the Gross-Pitaevskki description are not able to reach a strictly zero propagation speed for the lowest mode, beyond Mean Field descriptions have succeeded in observing rigorously the supersolid to normal solid transition~\cite{Saccani2012}.}}. As we shall see next, this transition corresponds to the breaking of quantum phase coherence.

To further explore the nature of the instabilities we compute the second-order correlations of the fluctuation fields. At the lower density boundary, when the QC amplitude is small and the ground state wave function is dominated by the main Fourier amplitude of the pattern ($c_1$), we find that $Z$, $Z_\perp$ and all elastic couplings in $\tilde{C}$ approaches zero as $c_1^2$, while $\gamma$ and $\gamma_\perp$ diverges as $c_1^{-2}$. This behavior lead us to the conclusion that in this regime $\langle\delta n_0^2\rangle$ and $\langle \theta^2\rangle$ remain finite, while $\langle \bm{\Pi}^2\rangle\propto c_1^2$, $\langle\bm{\Pi}_\perp^2\rangle\propto c_1^2$, $\langle\textbf{u}^2\rangle\propto c_1^{-2}$  and $\langle\textbf{w}^2\rangle\propto c_1^{-2}$, as $c_1\rightarrow0$, with $c_1\gg c_{n>1}$. In this scenario the average wave function can be roughly approximated as $\langle\phi(x,\tau)\rangle\approx\sqrt{\rho}\sum_\textbf{n}c_\textbf{n}\exp(i\textbf{G}_\textbf{n}\cdot\textbf{x})\times\exp(-G_\textbf{n}^2\langle\textbf{u}^2\rangle/4-G_{\textbf{n},\perp}^2\langle\textbf{w}^2\rangle/4-\langle\theta^2\rangle/2)$. This result confirms that close enough to the superfluid-quantum-quasicrystal stability boundary, where we have large phonon and phason fluctuations, there is a strong non-perturbative ``renormalization" of the Fourier amplitudes of the quasicrystal profile~\footnote{The strong renormalization process is signaled by the non-analytic behavior of $\langle\phi\rangle$ in the limit $c_1\rightarrow 0$ upon inclusion of phonon and phason fluctuations }. This phenomenology suggests that in cases in which the mean field superfluid to quantum quasicrystal phase transition is weak enough, the nature of the transition could be modified to a continuous one, even with the presence of intermediate hexatic-like phases. 

At the upper boundary of the quantum quasicrystal phase ($\gamma \rho\rightarrow 1^+$) all couplings of the effective action in Eq. \eqref{action} remain finite, however in this limit the propagation speed of the lower hybridized mode approaches zero, as observed  before. This behavior  signals the disappearance of this mode, in the same way in which the lower energy excitation mode disappears at the supersolid to normal solid phase transition~\cite{Saccani2012}. Calculating the correlations of the fluctuation fields in this regime, we obtain to leading order that $\langle\delta n_0^2\rangle$ $\langle\textbf{u}^2\rangle$, $\langle\textbf{w}^2\rangle$ and $\langle\bm{\Pi}_\perp^2\rangle$ remain finite as the stability boundary is approached, while $\langle \theta^2\rangle\propto(\gamma\rho-1)^{-1/2} $ and $\langle\bm{\Pi}^2\rangle\propto(\gamma\rho-1)^{-1/2}$. This result is to be expected since in the high $\rho U$ regime the localization of particles should eventually lead the system to the loss of quantum phase coherence. Interestingly, due to the effective $\bm{\Pi}\cdot\bm{\nabla}\theta$ interaction the de-stabilization of the $\theta(\textbf{x},\tau)$ field also produces the destabilization of the $\bm{\Pi}(\textbf{x},\tau)$ field, indicating the natural breakdown of our theory at this phase boundary. All evidence available in this case indicates a second order phase transition in this limit, as observed in supersolid systems~\cite{Macri2013,Cinti_2014,Cinti_2017,Heinonen2019}.

\textit{Conclusions.}--We have developed the elastic theory of a self-organized quantum quasicrystal with 8, 10 or 12-fold rotational symmetry. We obtained the effective action of the system described in terms of the global wave function phase, the two-dimensional phonon and phason fields and their respective conjugate density fields. From here, we were able to prove from first principles the existence of five gapless extended low energy excitation modes~\footnote{It is worth to mention, that due to the lack of periodicity quantum quasicrystals can also host particle-hole localized excitations with an energy arbitrarily close to zero. See Refs.~\cite{SvBaPr2015,Zhu2024}.} for the considered two-dimensional QC structures ~\cite{Lifshitz2014,Sandbrink2014}. We showed that the presence of the $\bm{\Pi}\cdot\bm{\nabla}\theta$ coupling in our effective theory produces the hybridization of the condensate sound mode with phonon and phason modes, leading to a rich phenomenology depending on the QC structure. A particularly interesting finding was that, due to the phonon-phason coupling, the octagonal QC structure develops an anisotropic speed of sound that depends on direction of the propagation, as well as that all modes show a mixed longitudinal-transversal character. Finally, our results shed light on the behavior of the low energy fluctuation fields at the thermodynamic boundaries of the quantum quasicrystal phase, elucidating possible scenarios for phase transitions in two dimensions.            

\begin{acknowledgments}
{\it Acknowledgments:} A.M.C. acknowledge MPIPKS for financial support and hospitality.
\end{acknowledgments}

\bibliography{bibliography} 
\bibliographystyle{ieeetr}

\pagebreak
\widetext
\begin{center}
\textbf{\large Supplemental Materials: Low energy excitations in super quasicrystals: a generalized elasticity approach }
\end{center}

\setcounter{equation}{0}
\setcounter{figure}{0}
\setcounter{table}{0}
\setcounter{page}{1}
\makeatletter
\renewcommand{\theequation}{S\arabic{equation}}
\renewcommand{\thefigure}{S\arabic{figure}}
\renewcommand{\bibnumfmt}[1]{[S#1]}
\renewcommand{\citenumfont}[1]{S#1}

\section{I. Construction of the effective action}
In this section we present details about the  construction of the effective action for the low energy behavior of the octagonal, decagonal and dodecagonal cluster quasicrystalline structures. We assume that the pair interaction potential $v(\textbf{x})$ is such that the model self-stabilizes one of these phases as its ground state in the quantum regime of the system~\cite{MeZaDi2022}. By quantum regime, we refer to the region of parameters in which particle delocalization is strong enough to produce a finite superfluid fraction in the system. As a consequence, we consider that the stationary configuration that minimizes the action in Eq. (1) can be expressed as a Fourier expansion of the form
\begin{equation}
\phi_0(\textbf{x})=\sqrt{\rho}\sum_{\textbf{n}} c_\textbf{n}\exp(i \textbf{G}_{\textbf{n}}\cdot\textbf{x}),
    \label{gE}
\end{equation}
where $c_\textbf{n}$ and $\textbf{G}_\textbf{n}$ represent the amplitudes and wave vectors of the Fourier expansion of $\phi_0$. Moreover, for the kind of QCs considered in this work we have $\textbf{G}_{\textbf{n}}=k(n_1\textbf{e}_1+n_2\textbf{e}_2+n_3\textbf{e}_3+n_4\textbf{e}_4)$, where $k$ stands for the characteristic wave vector of the quasicrystal and the basis vectors $\textbf{e}_i$'s are specific for each kind of pattern. The sum in the Fourier expansion is performed over a four dimensional integer vector $\textbf{n}$ whose components $n_i$'s are all integer numbers. A possible definition of the vectors $\textbf{e}_i$'s for each of the cases considered is presented in Table \ref{tab1}. 
\begin{table}[h!]
\centering
\begin{tabular}{|l|l|l|l|l|}
\hline
 QC rot. sym. & \hspace{0.8cm} $\textbf{e}_1$ & \hspace{1.3cm} $\textbf{e}_2$ & \hspace{1.3cm} $\textbf{e}_3$ & \hspace{1.3cm} $\textbf{e}_4$ \\ \hline
\hspace{0.4cm} 8-fold & $(\cos(0),\sin(0))$ & $(\cos(\pi/4),\sin(\pi/4))$ & $(\cos(\pi/2),\sin(\pi/2))$ & $(\cos(3\pi/4),\sin(3\pi/4))$ \\ \hline
\hspace{0.4cm} 10-fold & $(\cos(0),\sin(0))$ & $(\cos(2\pi/5),\sin(2\pi/5))$ & $(\cos(4\pi/5),\sin(4\pi/5))$ & $(\cos(6\pi/5),\sin(6\pi/5))$ \\ \hline
\hspace{0.4cm} 12-fold & $(\cos(0),\sin(0))$ & $(\cos(2\pi/3),\sin(2\pi/3))$ & $(\cos(\pi/6),\sin(\pi/6))$ & $(\cos(3\pi/2),\sin(3\pi/2))$  \\ \hline
\end{tabular}
\caption{Selection of the basis $\{\textbf{G}_{\textbf{n}}\}$ for each quasi-periodic structure considered.}
\label{tab1}
\end{table}

The low energy excitation properties can be accessed considering the effects of long wave-length fluctuations in the Fourier amplitudes and in its corresponding conjugate phases, in the same spirit as presented in ref \cite{Pretko2018}. In our case we propose a general ansatz for the perturbed wave function field of the form 
\begin{eqnarray}
    &&\nonumber\phi(\textbf{x},\tau)=\sqrt{\rho}c_0\sqrt{1+\frac{\delta n_0(\textbf{x},\tau)}{\rho c_0^2}}\exp(i\theta(\textbf{x},\tau))
    +\sum_{\textbf{n}\neq0}\sqrt{\rho}c_\textbf{n}\sqrt{1+\frac{\textbf{G}_{\textbf{n}}\cdot\bm{\Pi}(\textbf{x},\tau)}{Z}+\frac{\textbf{G}_{\textbf{n},\perp}\cdot\bm{\Pi}_{\perp}(\textbf{x},\tau)}{Z_{\perp}}}\\ 
    &&\exp\left(i \textbf{G}_{\textbf{n}}\cdot\textbf{x}+i \textbf{G}_{\textbf{n}}\cdot\textbf{u}(\textbf{x},\tau)
    + i \textbf{G}_{\textbf{n},\perp}\cdot\textbf{w}(\textbf{x},\tau)
    +i\theta(\textbf{x},\tau)\right),
\end{eqnarray}
where the set $\{\textbf{G}_{\textbf{n},\perp}\}$ is constructed as a permutation of the set $\{\textbf{G}_{\textbf{n}}\}$ such that the extended four dimensional basis $\tilde{\textbf{G}}_{\textbf{n}}=\textbf{G}_{\textbf{n}}\oplus\textbf{G}_{\textbf{n},\perp}$, is orthogonal in the sense that $\sum_{\textbf{n}}\tilde{\textbf{G}}_{\textbf{n},\alpha}\tilde{\textbf{G}}_{\textbf{n},\beta}\propto\delta_{\alpha,\beta}$~\cite{Levine1984,Levine1985,Socolar1989,Ding1993}. For the purpose of calculations we can write $\textbf{G}_{\textbf{n},\perp}=k(n_1\textbf{e}_{1,\perp}+n_2\textbf{e}_{2,\perp}+n_3\textbf{e}_{3,\perp}+n_4\textbf{e}_{4,\perp})$, where the basis $\{\textbf{e}_{i,\perp}\}$ corresponding to our particular choice of $\{\textbf{e}_{i}\}$ is chosen as shown in Table \ref{tab2}. Moreover is important to notice that the index vector $\{n_i\}$ for two corresponding $\textbf{G}_{\textbf{n}}$ and $\textbf{G}_{\textbf{n},\perp}$ coincide. 

\begin{table}[h!]
\centering
\begin{tabular}{|l|l|l|l|l|}
\hline
 QC rot. sym. & \hspace{0.8cm} $\textbf{e}_{1,\perp}$ & \hspace{1.3cm} $\textbf{e}_{2,\perp}$ & \hspace{1.3cm} $\textbf{e}_{3,\perp}$ & \hspace{1.3cm} $\textbf{e}_{4,\perp}$ \\ \hline
\hspace{0.4cm} 8-fold & $(\cos(0),\sin(0))$ & $(\cos(3\pi/4),\sin(3\pi/4))$ & $(\cos(3\pi/2),\sin(3\pi/2))$ & $(\cos(\pi/4),\sin(\pi/4))$ \\ \hline
\hspace{0.4cm} 10-fold & $(\cos(0),\sin(0))$ & $(\cos(6\pi/5),\sin(6\pi/5))$ & $(\cos(2\pi/5),\sin(2\pi/5))$ & $(\cos(8\pi/5),\sin(8\pi/5))$ \\ \hline
\hspace{0.4cm} 12-fold & $(\cos(0),\sin(0))$ & $(\cos(4\pi/3),\sin(4\pi/3))$ & $(\cos(5\pi/6),\sin(5\pi/6))$ & $(\cos(3\pi/2),\sin(3\pi/2))$  \\ \hline
\end{tabular}
\caption{Selection of the basis $\{\textbf{G}_{\textbf{n},\perp}\}$ for each quasi-periodic structure considered.}
\label{tab2}
\end{table}
Since we are interested in the low energy behavior of the system we can work in the hydrodynamic regime, in which the fluctuation fields introduced vary slowly in space and imaginary time. Mathematically, this argument allows us to neglect all components of the effective theory containing the product of slowly varying terms with spatially oscillatory contributions.

Now we can proceed with the evaluation of each term of the original action (Eq. \eqref{original_action}) as a function of the fluctuation fields, keeping only up to quadratic contributions. The first contribution can be recast in the symmetric form 
\begin{equation}
\delta S_1=\int d\tau d^2x\frac{1}{2}\left(\phi(\textbf{x},\tau)^*\partial_\tau\phi(\textbf{x},\tau)-\phi(\textbf{x},\tau)\partial_\tau\phi(\textbf{x},\tau)^*\right),
\end{equation}
which up to quadratic order in the fluctuation fields yields
\begin{equation}
\delta S_1=\int d\tau d^2x \left(i \delta n_0(\textbf{x},\tau)\partial_\tau\theta(\textbf{x},\tau)+i \left(\frac{\rho\sum_\textbf{n}c_\textbf{n}^2\textbf{G}_\textbf{n}^2}{2Z}\right)\bm{\Pi}(\textbf{x},\tau)\cdot\partial_\tau\textbf{u}(\textbf{x},\tau)+i \left(\frac{\rho\sum_\textbf{n}c_\textbf{n}^2\textbf{G}_{\textbf{n},\perp}^2}{2Z_\perp}\right)\bm{\Pi}_\perp(\textbf{x},\tau)\cdot\partial_\tau\textbf{w}(\textbf{x},\tau)\right).
\end{equation}
In this way, by choosing $Z=\rho\sum_\textbf{n}c_\textbf{n}^2\textbf{G}_\textbf{n}^2/2$ and $Z_\perp=\rho\sum_\textbf{n}c_\textbf{n}^2\textbf{G}_{\textbf{n},\perp}^2/2$ we obtain for $\delta S_1$, the expected structure combining the phases and the corresponding conjugate density fluctuation fields. 

The second contribution of our action can be recast as
\begin{equation}
\delta S_2=\int d\tau d^2x\frac{1}{2}\bm{\nabla}\phi(\textbf{x},\tau)^*\cdot\bm{\nabla}\phi(\textbf{x},\tau).
\end{equation}
In this case, the use of our ansatz leads us to the expression 
\begin{eqnarray}
\nonumber
    \delta S_2&=&\delta S_2^0+\int d\tau d^2x\left\{ \frac{1}{8\rho c_0^2}(\bm{\nabla} \delta n_0)^2+\frac{1}{8Z}\left((\bm{\nabla} \Pi_x)^2+(\bm{\nabla} \Pi_y)^2\right)+\frac{1}{8Z_\perp}\left((\bm{\nabla} \Pi_{x,\perp})^2+(\bm{\nabla} \Pi_{y,\perp})^2\right)\right.\\
    &+&\left.\frac{\rho c_0^2}{2}(\bm{\nabla}\theta)^2+\frac{\rho\sum_{\textbf{n}\neq0}c_\textbf{n}^2}{2}(\bm{\nabla}\theta)^2+\frac{Z}{2}\left((\bm{\nabla} u_x)^2+(\bm{\nabla} u_y)^2\right)+\frac{Z_\perp}{2}\left((\bm{\nabla} w_x)^2+(\bm{\nabla}w_y)^2\right)+\bm{\Pi}\cdot\bm{\nabla}\theta \right\},
    \label{eq3}
\end{eqnarray}
where $\delta S_2^0$, represents a fluctuation-independent contribution. The linear order contributions in the fluctuation fields are not presented in this and the rest of the terms of the effective action because they have a zero total contribution, once we are considering fluctuations around the global minimum of our action. At this point it is worth noticing that the particular form considered for our ansatz in Eq. \eqref{Eqflut} as well as the appropriate selection of $Z$ and $Z_\perp$ are responsible for producing a form of $\delta S_1$ and $\delta S_2$ preserving the expected relation between the couplings associated to conjugate density and phase fields. Other choices for this ansatz, for instance including  a dependence of $Z$ and $Z_\perp$ with $\vert\textbf{G}_\textbf{n}\vert$, even when preserves the symmetries of the original expression will not be able to reproduce this property.  

The third contribution of the original microscopic action, $\delta S_3=\int d\tau d^2x\mu\phi(\textbf{x},\tau)^*\phi(\textbf{x},\tau)$, does not provide any quadratic contribution to the total effective action in the long wave limit. Finally, we can proceed with the contribution resulting from the non-local interaction term 
\begin{equation}
\delta S_4=\int d\tau d^2xd^2x'\frac{1}{2}U v(\textbf{x}-\textbf{x}')\vert\phi(\textbf{x},\tau)\vert^2\vert\phi(\textbf{x}',\tau)\vert^2.
\end{equation}
This last term produces two kinds of contributions, one including only density-density interactions defined as $\delta S_{4,1}$ and a second one corresponding to the phonons and phasons deformation fields defined as $\delta S_{4,2}$. Terms of density-phase interactions cancel out exactly due to the reflection and rotational symmetries of the modulated patterns considered. Taking into account that in the determination of $\delta S_{4,1}$ and $\delta S_{4,2}$ we have to deal with a convolution with a non-local interaction kernel, it is easier to perform the evaluation of these terms in momentum space. After a lengthy but strightforward calculation we obtained that $\delta S_{4,1}$ in the long wave limit is given as
\begin{eqnarray}
\delta S_{4,1}&=&\int d\tau \frac{d^2q}{(2\pi)^2}\frac{1}{2}\left(\gamma_0\delta \hat{n}_0(\textbf{q},\tau)\delta \hat{n}_0(-\textbf{q},\tau)+\gamma\hat{\bm{\Pi}}(\textbf{q},\tau)\cdot\hat{\bm{\Pi}}(-\textbf{q},\tau)+\gamma_\perp\hat{\bm{\Pi}}_\perp(\textbf{q},\tau)\cdot\hat{\bm{\Pi}}_\perp(-\textbf{q},\tau)
\right),
\end{eqnarray}
where 
\begin{eqnarray}
\nonumber
\gamma_0&=&-\frac{1}{2}\sum_{\textbf{n1}\neq0,\textbf{n3},\textbf{n4}}U\hat{v}(\textbf{G}_{\textbf{n1}})\frac{c_\textbf{n1}c_\textbf{n3}c_\textbf{n4}}{c_0^{3}}\delta(\textbf{G}_{\textbf{n1}}+\textbf{G}_{\textbf{n3}}-\textbf{G}_{\textbf{n4}},\textbf{0})+\sum_\textbf{n}U\hat{v}(\textbf{G}_\textbf{n})\frac{c_\textbf{n}^2}{c_0^2},\\ \nonumber
\gamma&=&-\frac{1}{4}\sum_{\textbf{n1},\textbf{n2},\textbf{n3},\textbf{n4}}U\hat{v}(\textbf{G}_{\textbf{n1}}-\textbf{G}_{\textbf{n2}})\rho^2\frac{\Pi_{i=1}^4 c_\textbf{ni}}{Z^2}(\textbf{G}_{\textbf{n1}}-\textbf{G}_{\textbf{n2}})_x^2\delta(\textbf{G}_{\textbf{n1}}-\textbf{G}_{\textbf{n2}}+\textbf{G}_{\textbf{n3}}-\textbf{G}_{\textbf{n4}},\textbf{0}),\\
\gamma_\perp&=&-\frac{1}{4}\sum_{\textbf{n1},\textbf{n2},\textbf{n3},\textbf{n4}}U\hat{v}(\textbf{G}_{\textbf{n1}}-\textbf{G}_{\textbf{n2}})\rho^2\frac{\Pi_{i=1}^4 c_\textbf{ni}}{Z_{\perp}^2}(\textbf{G}_{\textbf{n1},\perp}-\textbf{G}_{\textbf{n2},\perp})_x^2\delta(\textbf{G}_{\textbf{n1}}-\textbf{G}_{\textbf{n2}}+\textbf{G}_{\textbf{n3}}-\textbf{G}_{\textbf{n4}},\textbf{0}).
\end{eqnarray}
The obtained expression for each $\gamma$ reduces to the one well known for the homogeneous scenario ($c_0=1$ and $c_{\textbf{n}\neq0}=0$), $\gamma_0=U\hat{v}(0)$, $\gamma=0$ and $\gamma_\perp=0$. It is also important to remark that since $\hat{V}(\textbf{k})$ has a negative part responsible for exciting the main Fourier modes of the modulated pattern, we will have $\gamma_0$, $\gamma$ and $\gamma_\perp$ as positive parameters, which guarantees the stability of our effective theory. We should note that in the long wave limit the terms proportional to the gradient of the density fluctuation fields in $\delta S_2$ are subleading respect to the contribution in $\delta S_{4,1}$, which is why these terms do not appear in Eq. \eqref{action} of the main text.     

The last contribution in $\delta S_4$ corresponds to the phonon and phason interaction. This term can be written in general as 
\begin{eqnarray}
\nonumber
\delta S_{4,2}&=&\int d\tau\frac{d^2q}{(2\pi)^2}\frac{U \rho^2}{4}\sum_{\textbf{n}}\Pi_{i=1}^4 c_{\textbf{ni}}~\delta(\textbf{G}_{\textbf{n1}}-\textbf{G}_{\textbf{n2}}+\textbf{G}_{\textbf{n3}}-\textbf{G}_{\textbf{n4}},\textbf{0})\\ \nonumber
&\times&\left(2\hat{v}(\textbf{q}-(\textbf{G}_{\textbf{n1}}-\textbf{G}_{\textbf{n2}}))-2\hat{v}(\textbf{G}_{\textbf{n1}}-\textbf{G}_{\textbf{n2}})\right)\left((\textbf{G}_{\textbf{n1}}-\textbf{G}_{\textbf{n2}})\cdot\hat{\textbf{u}}(\textbf{q},\tau)+(\textbf{G}_{\textbf{n1},\perp}-\textbf{G}_{\textbf{n2},\perp})\cdot\hat{\textbf{w}}(\textbf{q},\tau) \right)\\
&&\left((\textbf{G}_{\textbf{n1}}-\textbf{G}_{\textbf{n2}})\cdot\hat{\textbf{u}}(-\textbf{q},\tau)+(\textbf{G}_{\textbf{n1},\perp}-\textbf{G}_{\textbf{n2},\perp})\cdot\hat{\textbf{w}}(-\textbf{q},\tau) \right).
\label{eq1}
\end{eqnarray}
It is possible to convince ourselves that unlike all previous terms in our effective action this contribution has a different form for each quasicrystal. As expected in the long wave limit $\delta S_{4,2}$ reproduces the structure of the well-known classical elastic Hamiltonian for each kind of quasicrystal structure~\cite{Levine1984,Levine1985,Socolar1989,Ding1993,Fan2016} yielding 
\begin{eqnarray}
\nonumber
\delta S_{4,2}&=&\frac{1}{2}\int d\tau\frac{d^2q}{(2\pi)^2}\left[((2\mu+\lambda)q_x^2+\mu q_y^2)\hat{u}_x(\textbf{q},\tau)\hat{u}_x(-\textbf{q},\tau)+((2\mu+\lambda)q_y^2+\mu q_x^2)\hat{u}_y(\textbf{q},\tau)\hat{u}_y(-\textbf{q},\tau)\right.\\
\nonumber
&+&\left. (2\mu+2\lambda)q_xq_y\hat{u}_x(\textbf{q},\tau)\hat{u}_y(-\textbf{q},\tau)+(K_1q_x^2+(K_1+K_2+K_3) q_y^2)\hat{w}_x(\textbf{q},\tau)\hat{w}_x(-\textbf{q},\tau)\right.\\ \nonumber
&+&\left. (K_1q_y^2+(K_1+K_2+K_3) q_x^2)\hat{w}_y(\textbf{q},\tau)\hat{w}_y(-\textbf{q},\tau)+2(K_2+K_3)q_xq_y\hat{w}_x(\textbf{q},\tau)\hat{w}_y(-\textbf{q},\tau)\right.\\ \nonumber
&+&\left.(R q_x^2-R q_y^2)(\hat{u}_x(\textbf{q},\tau)\hat{w}_x(-\textbf{q},\tau)+\hat{u}_y(\textbf{q},\tau)\hat{w}_y(-\textbf{q},\tau))+2Rq_xq_y(\hat{u}_x(\textbf{q},\tau)\hat{w}_y(-\textbf{q},\tau)-\hat{u}_y(\textbf{q},\tau)\hat{w}_x(-\textbf{q},\tau))
\right].\\
\label{eq2}
\end{eqnarray}
The expression presented corresponds to the most general scenario, consistent with the octagonal quasi-crystalline pattern. In the case of the decagonal quasi-crystal we will have $K_2=-K_3$, which makes zero the off-diagonal phason-phason interaction proportional to $\hat{w}_x(\textbf{q},\tau)\hat{w}_y(-\textbf{q},\tau)$ and finally, in the case of the dodecagonal quasicrystal we will have $R=0$, which makes zero all elastic phonon-phason interactions. It is important to mention that the particular form obtained for our elastic phonon-phason action is a direct consequence of the particular choice we made for $\textbf{G}_{\textbf{n},\perp}$. Different but equivalent forms can be obtained for $\delta S_{4,2}$ if we consider other permutations of $\textbf{G}_\textbf{n}$ in the definition of $\textbf{G}_{\textbf{n},\perp}$. Furthermore, from the comparison between Eqs. \ref{eq1} and \ref{eq2} it is possible to conclude that

\begin{align}
(2\mu+\lambda)&=\frac{U \rho^2}{2}\sum_{\textbf{n}}\Pi_{i=1}^4 c_{\textbf{ni}}~\delta(\textbf{G}_{\textbf{n1}}-\textbf{G}_{\textbf{n2}}+\textbf{G}_{\textbf{n3}}-\textbf{G}_{\textbf{n4}},\textbf{0})(\textbf{G}_{\textbf{n1}}-\textbf{G}_{\textbf{n2}})_x^2\partial_{qx}^2\hat{v}(\textbf{q}-(\textbf{G}_{\textbf{n1}}-\textbf{G}_{\textbf{n2}}))\vert_{\textbf{q}=\textbf{0}}\\
\mu&=\frac{U \rho^2}{2}\sum_{\textbf{n}}\Pi_{i=1}^4 c_{\textbf{ni}}~\delta(\textbf{G}_{\textbf{n1}}-\textbf{G}_{\textbf{n2}}+\textbf{G}_{\textbf{n3}}-\textbf{G}_{\textbf{n4}},\textbf{0})(\textbf{G}_{\textbf{n1}}-\textbf{G}_{\textbf{n2}})_x^2\partial_{qy}^2\hat{v}(\textbf{q}-(\textbf{G}_{\textbf{n1}}-\textbf{G}_{\textbf{n2}}))\vert_{\textbf{q}=\textbf{0}}\\
K_1&=\frac{U \rho^2}{2}\sum_{\textbf{n}}\Pi_{i=1}^4 c_{\textbf{ni}}~\delta(\textbf{G}_{\textbf{n1}}-\textbf{G}_{\textbf{n2}}+\textbf{G}_{\textbf{n3}}-\textbf{G}_{\textbf{n4}},\textbf{0})(\textbf{G}_{\textbf{n1},\perp}-\textbf{G}_{\textbf{n2},\perp})_x^2\partial_{qx}^2\hat{v}(\textbf{q}-(\textbf{G}_{\textbf{n1}}-\textbf{G}_{\textbf{n2}}))\vert_{\textbf{q}=\textbf{0}}\\
K_T&=\frac{U \rho^2}{2}\sum_{\textbf{n}}\Pi_{i=1}^4 c_{\textbf{ni}}~\delta(\textbf{G}_{\textbf{n1}}-\textbf{G}_{\textbf{n2}}+\textbf{G}_{\textbf{n3}}-\textbf{G}_{\textbf{n4}},\textbf{0})(\textbf{G}_{\textbf{n1},\perp}-\textbf{G}_{\textbf{n2},\perp})_x^2\partial_{qy}^2\hat{v}(\textbf{q}-(\textbf{G}_{\textbf{n1}}-\textbf{G}_{\textbf{n2}}))\vert_{\textbf{q}=\textbf{0}}\\ \nonumber
R&=U \rho^2\sum_{\textbf{n}}\Pi_{i=1}^4 c_{\textbf{ni}}~\delta(\textbf{G}_{\textbf{n1}}-\textbf{G}_{\textbf{n2}}+\textbf{G}_{\textbf{n3}}-\textbf{G}_{\textbf{n4}},\textbf{0})(\textbf{G}_{\textbf{n1},\perp}-\textbf{G}_{\textbf{n2},\perp})_x(\textbf{G}_{\textbf{n1},\perp}-\textbf{G}_{\textbf{n2},\perp})_x\\
\times&\partial_{qx}^2\hat{v}(\textbf{q}-(\textbf{G}_{\textbf{n1}}-\textbf{G}_{\textbf{n2}}))\vert_{\textbf{q}=\textbf{0}},
\end{align}
where $K_T=K_1+K_2+K_3$. These expressions complete the determination of all couplings in our effective theory from the knowledge of the ground state wave function. Finally, we can notice that the gradient terms of the phonon and phason fields in Eq. \eqref{eq3} can be absorbed in $\delta S_{4,2}$ modifying the elastic coefficients as follows $(2\tilde{\mu}+\tilde{\lambda})=(2\mu+\lambda)+Z$, $\tilde{\mu}=\mu+Z$, $\tilde{K}_1=K_1+Z_\perp$ and $\tilde{K}_T=K+Z_\perp$. This allows us to write our total effective action in the following more compact way 
\begin{align}
    \nonumber
\delta S&=\int d\tau d^2x \left(i \delta n_0(\textbf{x},\tau)\partial_\tau\theta(\textbf{x},\tau)+i \bm{\Pi}(\textbf{x},\tau)\cdot\partial_\tau\textbf{u}(\textbf{x},\tau)+i \bm{\Pi}_\perp(\textbf{x},\tau)\cdot\partial_\tau\textbf{w}(\textbf{x},\tau)\right)+ \int d\tau d^2x\ \frac{1}{2}\left[\rho(\bm{\nabla}\theta)^2\right.\\ \nonumber
&\left.+\gamma_0\delta \hat{n}_0(\textbf{x},\tau)^2+\gamma\bm{\Pi}(\textbf{x},\tau)^2+\gamma_\perp\bm{\Pi}_\perp(\textbf{x},\tau)^2+2\bm{\Pi}\cdot\bm{\nabla}\theta+\textbf{E}^T~\tilde{C}~\textbf{E}\right],
\end{align}
where $\textbf{E}$ stands for the extended phonon-phason strain field defined as $\textbf{E}=\{\partial_xu_x,\partial_yu_y,(\partial_xu_y+\partial_yu_x)/2,(\partial_yu_x+\partial_xu_y)/2,\partial_xw_x,\partial_yw_y,\partial_yw_x,\partial_xw_y\}$ and $\tilde{C}$ represents the phonon-phason elastic tensor~\cite{Ding1993} given by

\begin{gather}
\nonumber
\tilde{C}=\\
\begin{bmatrix}
   \tilde{\lambda} +2\tilde{\mu} &\tilde{\lambda} & 0 & 0 & R & R & 0 & 0 \\
   \tilde{\lambda} &\tilde{\lambda}+2\tilde{\mu} & 0 & 0 & -R & -R & 0 & 0\\   
   0 & 0 & \tilde{\mu} & \tilde{\mu} & 0 & 0 & -R & R  \\
   0 & 0 & \tilde{\mu} & \tilde{\mu} & 0 & 0 & -R & R  \\
   R & -R & 0 & 0 & \tilde{K_1} & K_2 & 0 & 0 \\ 
   R & -R & 0 & 0 & K_2 & \tilde{K_1} & 0 & 0\\
   0 & 0 & -R & -R & 0 & 0 & \tilde{K}_T & K_3\\
   0 & 0 & R & R & 0 & 0 & K_3 & \tilde{K}_T    
   \end{bmatrix}.
\end{gather}

\section{II. Excitation properties of the dodecagonal quasicrystal}
As we can notice, the effective action in Eq. \eqref{action} can be written in the general form
\begin{equation}
\delta S=\frac{1}{2}\int \frac{d\omega}{(2\pi)}\frac{d^2q}{(2\pi)^2}\hat{\bm{\eta}}(\textbf{q},\omega)^\dagger \textbf{M}(\textbf{q},\omega)\hat{\bm{\eta}}(\vec{q},\omega),
\end{equation}
where the following definition of the Fourier transform of $f(\textbf{x},\tau)$ and its inverse have been used
\begin{eqnarray}
\hat{f}(\textbf{q},\omega)&=&\int\frac{d^2q d\omega}{(2\pi)^{3}}e^{i(\textbf{q}\cdot\textbf{x}-\omega\tau)}f(\textbf{x},\tau)\\
f(\textbf{x},\tau)&=&\int d^2x d\tau e^{-i(\textbf{q}\cdot\textbf{x}-\omega\tau)}\hat{f}(\textbf{q},\omega).
\end{eqnarray}
Moreover, $\hat{\bm{\eta}}(\textbf{q},\omega)=\{\delta \hat{n}_0(\textbf{q},\omega),\hat{\theta}(\textbf{q},\omega),\hat{\Pi}_x(\textbf{q},\omega),\hat{u}_x(\textbf{q},\omega),\hat{\Pi}_y(\textbf{q},\omega),\hat{u}_y(\textbf{q},\omega),\hat{\Pi}_{x,\perp}(\textbf{q},\omega),\hat{w}_x(\textbf{q},\omega),\hat{\Pi}_{y,\perp}(\textbf{q},\omega),\hat{w}_y(\textbf{q},\omega)\}$ and the interaction matrix $\textbf{M}(\textbf{q},\omega)$ is written as
\begin{gather}
\nonumber
\textbf{M}(\textbf{q},\omega)=
\begin{bmatrix}
   \gamma_0 &-\omega & 0 & 0 & 0 & 0 &0 &0 & 0 & 0 \\
   \omega & \rho q^2 & -iq_x & 0 & -iq_y & 0 & 0 &0 & 0 & 0 \\
   0 & iq_x & \gamma & -\omega & 0 & 0 &0 & 0 & 0 \\
   0 &0 & \omega & (2\tilde{\mu}+\tilde{\lambda})q_x^2+\tilde{\mu}q_y^2 & 0 & (\tilde{\mu}+\tilde{\lambda})q_xq_y &0 &\frac{R}{2}(q_x^2-q_y^2)& 0 & Rq_xq_y \\
   0 &i q_y & 0 & 0 & \gamma & -\omega & 0 &0 & 0 & 0 \\
   0 &0 & 0 & (\tilde{\mu}+\tilde{\lambda})q_xq_y & \omega &  (2\tilde{\mu}+\tilde{\lambda})q_y^2+\tilde{\mu}q_x^2 & 0 &-Rq_xq_y & 0 & \frac{R}{2}(q_x^2-q_y^2) \\
   0 & 0 & 0 & 0 & 0 & 0 & \gamma_\perp & -\omega & 0 & 0\\
   0 & 0 & 0 & \frac{R}{2}(q_x^2-q_y^2) & 0 & -Rq_xq_y & \omega & \tilde{K}_1q_x^2+\tilde{K}_Tq_y^2 & 0 & (K_2+K_3)q_xq_y\\
   0 & 0 & 0 & 0 & 0 & 0 & 0 & 0 & \gamma_\perp & -\omega \\
    0 & 0 & 0 & Rq_xq_y & 0 & \frac{R}{2}(q_x^2-q_y^2) & 0 & (K_2+K_3)q_x q_y & \omega & \tilde{K}_T q_x^2+\tilde{K}_1q_y^2
   \end{bmatrix}.
\end{gather}

As we can observe the interaction matrix $\mathbf{M}(\textbf{q},\omega)$ separates in two independent blocks of $6\times6$ and $4\times4$ when $R=0$, i.e. in the case of the dodecagonal quasicrystal pattern. This case corresponds to the simplest scenario, in which phasons and phonons fields do not interact with each other.

Now we can proceed with our study of the low energy behavior of the system. The dispersion relation $\omega(\textbf{q})$ corresponding to each excitation mode can be obtained by solving the equation $\det\left[\textbf{M}(\textbf{q},i\omega(\textbf{q}))\right]=0$. The solution of this equation leads to two groups of solutions corresponding to the two different blocks of the matrix $\textbf{M}(\textbf{q},\omega)$ in this case. The solutions arising from the $6\times6$ block of $\textbf{M}(\textbf{q},\omega)$ are given by
\begin{eqnarray}
\nonumber
\omega_{1}(\textbf{q})&=&\sqrt{A+\sqrt{A^2+B}}q\\ \nonumber
\omega_{2}(\textbf{q})&=&\sqrt{A-\sqrt{A^2+B}}q\\
\omega_{3}(\textbf{q})&=&\sqrt{ \gamma\tilde{\mu}}q,
\end{eqnarray}
where $A=(\rho\gamma_0+\gamma(2\tilde{\mu}+\tilde{\lambda}))/2$ and $B=(2\tilde{\mu}+\tilde{\lambda})\gamma_0(1-\gamma\rho)$. The form of the solutions indicates a hybridization process of the modes with dispersion relation $\omega_1(\textbf{q})$ and $\omega_2(\textbf{q})$, the analysis of eigenvectors associated to these eigenvalues allow us to conclude that these two excitation modes correspond to a hybridization of the longitudinal Bogoliubov mode with the longitudinal phonon mode, while $\omega_3(\textbf{q})$ corresponds to the transversal phonon mode. A similar procedure for the $4\times4$ block of $\textbf{M}(\textbf{q},\omega)$ yields the dispersion relations 
\begin{eqnarray}
\nonumber
\omega_{4}(\textbf{q})&=&\sqrt{\gamma_\perp \tilde{K}_1}q\\
\omega_{5}(\textbf{q})&=&\sqrt{ \gamma_\perp\tilde{K}_T}q,
\end{eqnarray}
corresponding to the longitudinal and transversal phason modes, respectively. 

\section{III. Fluctuation eigenstates of the dodecagonal quasicrystal}
The longitudinal or transversal character of the excitation modes in the dodecagonal quasicrystal refers to the way in which the vectorial fields $\bm{\Pi}(\textbf{x},\tau)$, $\textbf{u}(\textbf{x},\tau)$, $\bm{\Pi}_\perp(\textbf{x},\tau)$ and $\textbf{w}(\textbf{x},\tau)$ oscillate respect to the momentum of the excitation $\textbf{q}$. For a better understanding of the nature of these modes we calculated the corresponding eigenvectors $\hat{\bm{\eta}}_j(\textbf{q})$ in a situation in which $\textbf{q}$ is oriented along the "$x$"-axis, yielding 

\begin{eqnarray}
\nonumber
\bm{\hat{\eta}}_{1}(q_x)&=&\left \lbrace (c_1^2-\gamma(2\tilde{\mu}+\tilde{\lambda}))c_1 q_x,~-i\gamma_0(c_1^2-\gamma(2\tilde{\mu}+\tilde{\lambda})),~\gamma_0(2\tilde{\mu}+\tilde{\lambda})q_x,~-i\gamma_0c_1,0,0,0,0,0,0\right\rbrace\\ \nonumber
\bm{\hat{\eta}}_{2}(q_x)&=&\left \lbrace -(c_2^2-\gamma_0\rho)c_2 q_x,~i\gamma_0(c_2^2-\gamma_0\rho),~\gamma_0(2\tilde{\mu}+\tilde{\lambda})q_x,~-i\gamma_0c_2,0,0,0,0,0,0\right\rbrace\\ \nonumber
\bm{\hat{\eta}}_{3}(q_x)&=&\left \lbrace {0, 0, 0, 0, c_3 q_x, -i\gamma, 0, 0, 0, 0}\right\rbrace\\ \nonumber
\bm{\hat{\eta}}_{4}(q_x)&=&\left \lbrace {0, 0, 0, 0, 0, 0, c_4 q_x, -i \gamma_\perp, 0, 0}\right\rbrace\\ 
\bm{\hat{\eta}}_{5}(q_x)&=&\left \lbrace 0, 0, 0, 0, 0, 0, 0, 0, c_5 q_x, -i\gamma_\perp\right\rbrace,
\label{eigenvec}
\end{eqnarray}
where $c_i=\partial_q \epsilon_i(\textbf{q})$, refers to the sound velocity of the corresponding excitation mode. From here, the fluctuation fields in real space and time writes as $\bm{\eta}_j(\textbf{x},t)=\mathrm{Re}\left[\bm{\hat{\eta}}_j(\textbf{q})\exp(i(\textbf{q}\cdot\textbf{x}-\omega_jt))\right]$ and the correction to the ground state wave function as 
\begin{eqnarray}
\nonumber
\delta \psi&=&\sqrt{\rho}\left(\frac{\delta n_0(\textbf{x},t)}{2\rho c_0^2}+ic_0\theta(\textbf{x},t)+\frac{1}{2}\sum_{\textbf{n}\neq\textbf{0}}c_\textbf{n}\left(\frac{\textbf{G}_{\textbf{n}}\cdot\bm{\Pi}(\textbf{x},t)}{Z}+\frac{\textbf{G}_{\textbf{n},\perp}\cdot\bm{\Pi}_\perp(\textbf{x},t)}{Z}\right)\right.\\ \nonumber
&&\left.\times\exp(i\textbf{G}_{\textbf{n}}\cdot\textbf{x})+i\sum_{\textbf{n}\neq\textbf{0}}c_\textbf{n}\exp(i\textbf{G}_{\textbf{n}}\cdot\textbf{x})\times (\theta(\textbf{x},t)+\textbf{G}_\textbf{n}\cdot\textbf{u}(\textbf{x},t))+\textbf{G}_{\textbf{n},\perp}\cdot\textbf{w}(\textbf{x},t))\right.\biggr).
\end{eqnarray}
Comparing this expression with the usual ansatz for the solution of the linearized Gross-Pitaevskii equation~\cite{kunimi2012}, $ \delta\psi_{j,\textbf{q}}=u_{j,\textbf{q}}(\textbf{r})\exp(i(\textbf{q}\cdot\textbf{r}-\omega_j(\textbf{q})t))-v_{j,\textbf{q}}(\textbf{r})^*\exp(-i(\textbf{q}\cdot\textbf{r}-\omega_j(\textbf{q})t))$, we identify in each case the  functions $u_{j,\textbf{q}}(\textbf{r})$ and $v_{j,\textbf{q}}(\textbf{r})$. This allows us to compute the phase and density fluctuations for the excitation modes as $\Delta\rho_{j,\textbf{q}}=\vert u_{j,\textbf{q}}(\textbf{r})-v_{j,\textbf{q}}(\textbf{r})\vert^2$ and $\Delta\varphi_{j,\textbf{q}}=\vert u_{j,\textbf{q}}(\textbf{r})+v_{j,\textbf{q}}(\textbf{r})\vert^2$~\cite{Macri2013, Wu1996}. In Fig.\ref{fig1}(\textbf{c}-\textbf{l}) we show the behavior of these quantities considering typical parameters for the ground state wave function and for the pair interaction potential. For the numerical evaluation, the effective couplings of the theory can be roughly estimated using a single mode expansion for the wave function of the dodecagonal QC. In this case, we can write $\phi(\textbf{x})=\sqrt{\rho}(c_0+c_1\sum_{j=1}^{12}\exp(i k\textbf{e}_j\cdot\textbf{x}))$, where $\{\textbf{e}_i\}$ is a set of unitary vectors forming regular dodecagon. In the small $c_1$ limit we can reach to 
\begin{eqnarray}
Z&\approx&Z_\perp\approx6\rho k^2c_1^2\\
\gamma&\approx&\gamma_\perp\approx U\abs{\hat{v}(k)}/(6c_1^2k^2)\\
\gamma_0&\approx&U\hat{v}(0)\\
(2\tilde{\mu}+\tilde{\lambda})&\approx&\tilde{K}_T\approx6\rho k^2c_1^2+3U\rho^2c_1^2(k\hat{v}'(k)+3k^2\hat{v}''(k))\\
\tilde{\mu}&\approx&\tilde{K}_1\approx6\rho k^2c_1^2+3U\rho^2c_1^2(3 k\hat{v}'(k)+k^2\hat{v}''(k)).
\end{eqnarray}
Considering $U=1$, $\rho=0.4$, $k=1$,  $\hat{v}(0)=2$, $\hat{v}(k)=-1$, $\hat{v}'(k)=0$, $\hat{v}''(k)=1.5$, $c_0=1$, $c_1=0.15$ and a momentum of the excitation $\textbf{q}=0.1(1,0)$ we obtain the density and phase fluctuations patterns shown in Fig.1 \textbf{(c-l)}.

\section{IV. Numerical considerations for Fig. 2}
For the numerical evaluation associated to Fig.2 (a) we considered the following set of arbitrary parameters $\rho=4$, $\gamma_0=4$, $\gamma=1$ $\gamma_\perp=1$, $2\tilde{\mu}+\tilde{\lambda}=52$, $\tilde{\mu}=20$, $\tilde{K}_1=24$, $\tilde{K}_T=56$ and $R=32$.  

Moreover, as explained in the main text for Fig.2 (b) we need more realistic estimates of the behavior of the couplings of the effective theory in Eq.(3), these can be obtained from a single mode expansion, which for the decagonal quasicrystal, yields 
\begin{eqnarray}
Z&\approx&Z_\perp\approx5\rho k^2c_1^2\\
\gamma&\approx&\gamma_\perp\approx U\abs{\hat{v}(k)}/(5c_1^2k^2)\\
\gamma_0&\approx&U\hat{v}(0)\\
(2\tilde{\mu}+\tilde{\lambda})&\approx&5\rho k^2c_1^2+\frac{5}{2}U\rho^2c_1^2(k\hat{v}'(k)+3k^2\hat{v}''(k))\\
\tilde{\mu}&\approx&5\rho k^2c_1^2+\frac{5}{2}U\rho^2c_1^2(3 k\hat{v}'(k)+k^2\hat{v}''(k))\\
\tilde{K}_1=\tilde{K}_T&\approx&5\rho k^2c_1^2+5U\rho^2c_1^2(k\hat{v}'(k)+k^2\hat{v}''(k))\\
R&\approx&\frac{5}{2}U\rho^2c_1^2(-k\hat{v}(k)+k^2\hat{v}''(k)).
\end{eqnarray}
Additionally, for numerical evaluations we use the anzats  $c_1 (U\rho)=0.035+0.51(U\rho-0.2)/(U\rho)^{0.65}$. This ansatz is only intended to capture the two main features of the dominant Fourier amplitude of the QC pattern, i.e. $c_1(U\rho)$ is a strictly growing function defined from the QC phase boundary ($U\rho=0.2$). Secondly, $c_1(U\rho)$ should be concave reflecting the tendency of the system to avoid collapse as $U\rho$ is increased. Furthermore, for the pair interaction parameters we consider $\hat{v}(k)=-1$, $\hat{v}'(k)=0$, $\hat{v}''(k)=1$ and $\hat{v}(0)=1$.

\section{Behavior of the Density and Phase Local fluctuations in the Octagonal Quasicrystal Pattern}

In this section we present the typical behavior of the local density and phase fluctuations for the octagonal quasicrystal for different orientations $\alpha$ of the propagation direction of the excitation respect to one of the main directions of the quasicrystal. For numerical calculations all the couplings of our effective theory takes the same values considered for the construction of Fig.(\ref{fig2}). 

\begin{figure}[h!]
\centering
\includegraphics[width=\textwidth]{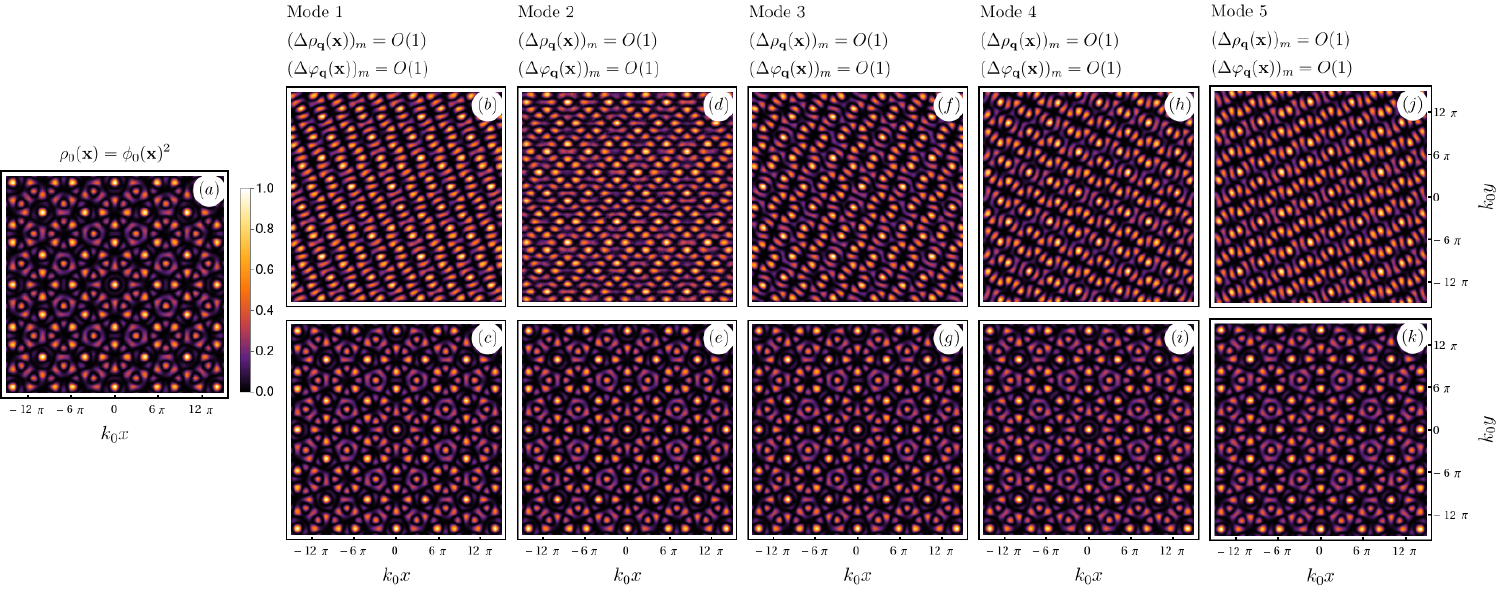}
        \caption{Typical density pattern \textbf{(a)} and low-energy excitations of a octagonal quantum quasicrystal (QQC) when the excitation propagates forming an angle $\alpha=\pi/16$ respect to one of the main directions of the quasicrystal \textbf{(b-k)}. The color scale included in \textbf{(a)} applies to all figures provided each profile is normalized by the maximum of the corresponding quantity. \textbf{(b-k)} Density fluctuations $\Delta \rho_\textbf{q}(\textbf{x})$ (first row) and phase fluctuations $\Delta \varphi_\textbf{q}(\textbf{x})$ (second row) in the low momentum regime for each gapless excitation mode according to its energy value ordered increasingly from left to right. The scaling behaviour of the density and phase fluctuations for each mode is given at the top of each column. See Ref.~\cite{sm} for detailed information. 
    }
    \label{fig3}
\end{figure}

\begin{figure}[ht]
\centering
\includegraphics[width=\textwidth]{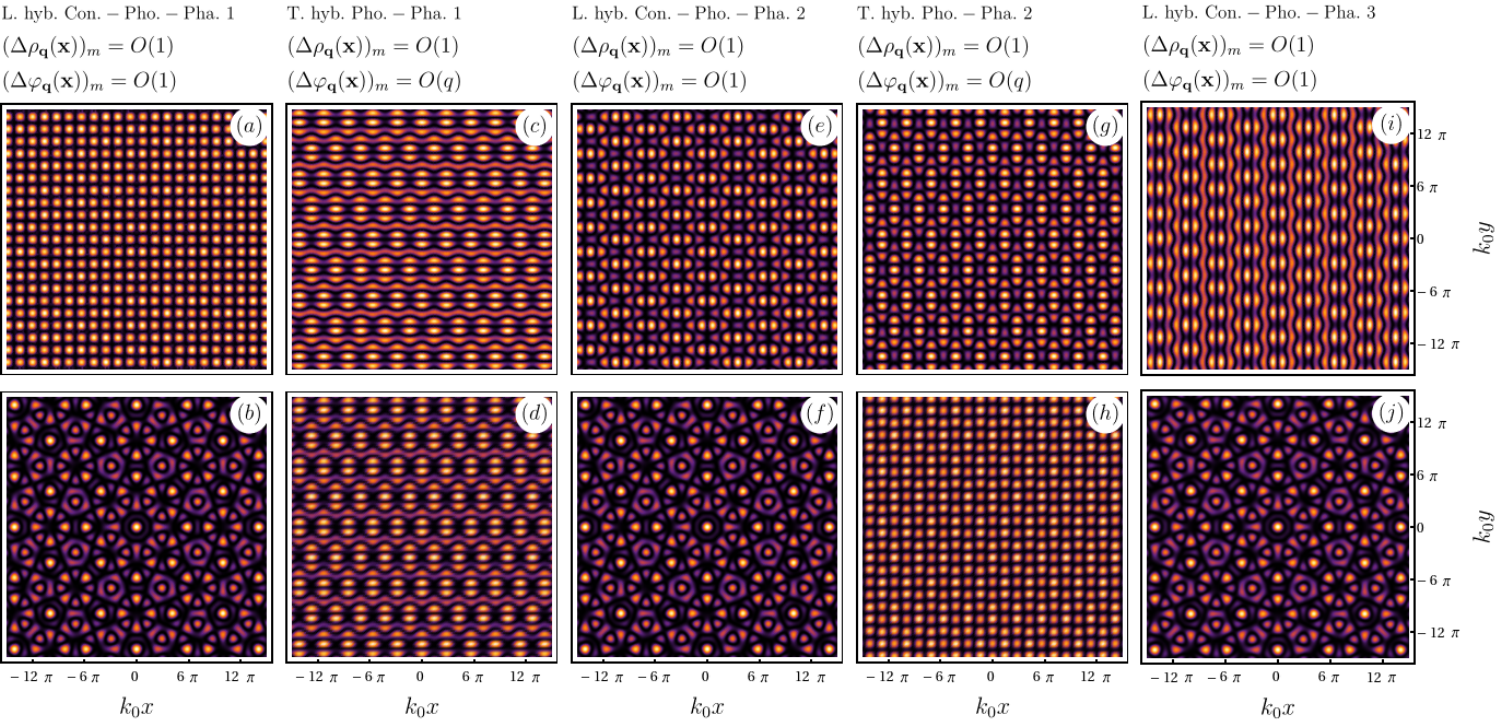}
        \caption{Typical low-energy Density fluctuations $\Delta \rho_\textbf{q}(\textbf{x})$ (first row) and phase fluctuations $\Delta \varphi_\textbf{q}(\textbf{x})$ (second row) in the low momentum regime for each gapless excitation in increasing order of the corresponding excitation energy value from left to right. The scaling behaviour of the density and phase fluctuations for each mode is given at the top of each column. The momentum of the excitation $\textbf{q}$ is chosen forming an angle $\alpha=0$ with one of the main directions of the quasicrystal, see Ref.~\cite{sm} for detailed information. 
    }
    \label{fig4}
\end{figure}

\begin{figure}[ht]
\centering
\includegraphics[width=\textwidth]{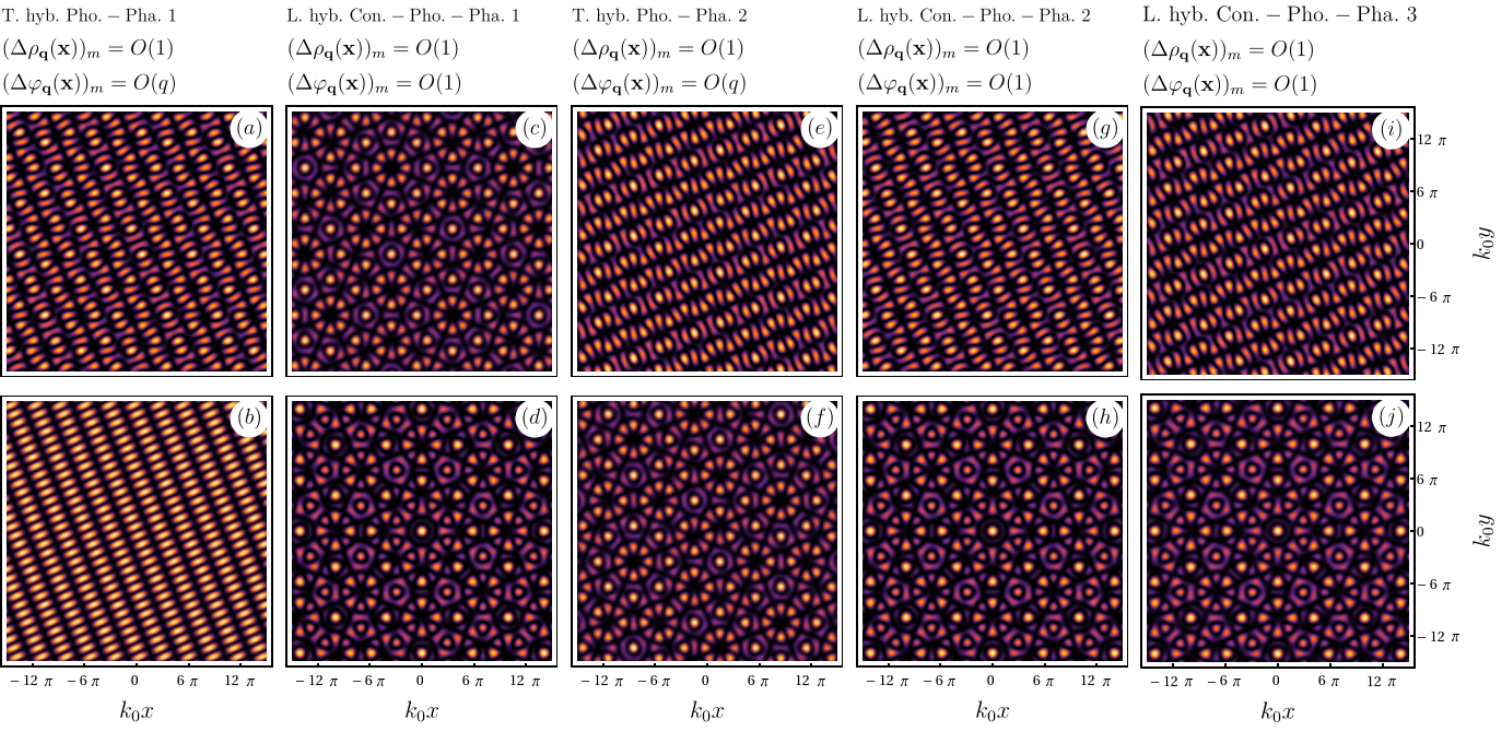}
        \caption{Typical low-energy Density fluctuations $\Delta \rho_\textbf{q}(\textbf{x})$ (first row) and phase fluctuations $\Delta \varphi_\textbf{q}(\textbf{x})$ (second row) in the low momentum regime for each gapless excitation in increasing order of the corresponding excitation energy value from left to right. The scaling behaviour of the density and phase fluctuations for each mode is given at the top of each column. The momentum of the excitation $\textbf{q}$ is chosen forming an angle $\alpha=\pi/8$ with one of the main directions of the quasicrystal, see Ref.~\cite{sm} for detailed information. 
    . 
    }
    \label{fig5}
\end{figure}

The first case presented corresponds to the most general scenario for which all low energy modes are hybridized. As a consequence the scaling of the density and phase fluctuations is $O(1)$ for all modes. For the numerical evaluation in this case we select $\alpha=\pi/16$, the results are presented in Fig.(\ref{fig3}). Next we consider separately the cases of $\alpha=0$ in Fig.(\ref{fig4}) and $\alpha=\pi/8$ in Fig.(\ref{fig5}). For these special directions, as they correspond to symmetry directions of the octagonal quasicrystal structure, modes are purely longitudinal or transversal and the hybridization takes place only between modes in each of these separate groups. Because of this feature the scaling of the density and phase fluctuations for the longitudinal and transversal modes remains different in these cases.

\end{document}